\documentclass[twoside,fleqn]{article}
\usepackage{espcrc2}
\usepackage[dvips]{epsfig}
\usepackage{latexsym}

\def\beq{\begin{equation}} \def\eeq{\end{equation}}
\def\lsim{\raise0.3ex\hbox{$<$\kern-0.75em\raise-1.1ex\hbox{$\sim$}}}
\def\gsim{\raise0.3ex\hbox{$>$\kern-0.75em\raise-1.1ex\hbox{$\sim$}}}
\def\H{{\scriptscriptstyle H}} 

\def\FPI{f_\pi}\def\fpi{\ifmmode\FPI\else$\FPI$\fi}
\def\MH{m_\H}\def\mh{\ifmmode\MH\else$\MH$\fi}
\def\LMH{\mu_\H}\def\lmh{\ifmmode\LMH\else$\LMH$\fi}
\def\FMH{\bar\mu_\H}\def\fmh{\ifmmode\FMH\else$\FMH$\fi}
\def\PPB{\langle\bar\psi\psi\rangle}\def\ppb{\ifmmode\PPB\else$\PPB$\fi}
\def\NT{N_\tau}\def\t{\ifmmode\NT\else$\NT$\fi}
\def\NZ{N_z}\def\z{\ifmmode\NZ\else$\NZ$\fi}
\def\TC{T_c}\def\tc{\ifmmode\TC\else$\TC$\fi}
\def\PQ{\langle P\rangle}\def\pq{\ifmmode\PQ\else$\PQ$\fi}
\newcommand{\AmS}{{\protect\the\textfont2
  A\kern-.1667em\lower.5ex\hbox{M}\kern-.125emS}}

\newcommand{\beqn} {\begin{equation}}
\newcommand{\eqn} {\end{equation}}

\newcommand{\slsh}[1] {#1\kern-.43em/}
\newcommand{\real}{{\sf I}\kern-.12em{\sf R}}
\newcommand{\comp}{{\sf I}\kern-.48em{\sf C}}
\newcommand{\nin} {\in\kern-.6em/}

\title{The Non-Perturbative ${\cal O}(g^6)$ Contribution
       to the Free Energy of Hot SU(N) Gauge Theory}

\author{M. L\"utgemeier
        \thanks{This work has been supported in parts by the Deutsche
                Forschungsgemeinschaft under contracts Pe 340/3-3 and
                340/6-2}
        with F. Karsch, A. Patk\`os, J. Rank\\
        Faculty of Physics, University of Bielefeld,
        P.O. Box 100131, 33501 Bielefeld, Germany\\
       }

\begin{document}

\begin{abstract}
The non-perturbative input necessary for the determination of 
the ${\cal O}(g^6)$
part of the weak coupling expansion of the free energy density for SU(2) and
SU(3) gauge theories is estimated. Although the perturbative information
completing the contribution to this order is missing, we give 
arguments that the magnetic fluctuations are dominated by
screened elementary magnetic gluons.
\end{abstract}

\maketitle

\section{Motivation \& Introduction}
Perturbation theory in its original formulation fails beyond ${\cal O}(g^5)$
in the calculation of the free energy density of non-Abelian gauge theories
\cite{Lin80}. Very recently Braaten \cite{Bra95} has pointed out that the
coefficients of the higher powers in the weak coupling expansion can
be determined by invoking non-perturbative information from the effective
theory of three-dimensional static magnetic fluctuations.

In Ref.~\cite{Bra95}
a systematic two-step separation of the perturbatively treatable fluctuations
from the static magnetic sector has been proposed. In the first step the full
(electric {\it and } magnetic) static sector is represented by an effective
three-dimensional theory. In this theory (called EQCD) a massive ($m_E$)
adjoint scalar field stands for the screened electric fluctuations.
In the second step this theory is matched onto an effective magnetic theory
(MQCD) with a separation cut-off $\Lambda_M$.

While the contribution of the non-static and static
electric modes to the free energy can be safely calculated perturbatively
(expansion parameters are $g(T)$ and $m_E/T$, respectively), the magnetic
sector is still non-perturbative and should be investigated by numerical
methods. This was done by simulating SU(3) pure gauge theory in 3 dimensions
which is just the MQCD.

We also calculated the string tension in this theory and compared it to
the 4 dimensional spatial string tension and to a possible magnetic mass.

\section{Part A: Free Energy}

According to the two step reduction program the high temperature free energy
of the full theory can be written as
\begin{equation}
f_{QCD}(T>>T_c)=f_{NS} + f_E + f_M.
\end{equation}

The first term contains all non-static diagrams in the full 4-d theory,
while the second term contains the contribution from EQCD.
It can be calculated savely in an expansion in $m_E/T$.

The non-perturbative information contained in the third term comes from
the minimal MQCD theory. The purpose of our paper is to present
first results of a lattice analysis of MQCD, ie. the 3-dimensional SU(N)
gauge theory. This does provide the non-perturbative input needed to
evaluate the ${\cal O}(g^6)$
part of $f_M$, denoted in the following by $f_M^{(0)}$.

With the Lagrangian $L_{MQCD}^{(0)} = \frac{1}{4} G_{ij}^a G_{ij}^a$
one has the following cut-off dependence:
\begin{eqnarray*}
  f_M^{(0)} &=& T\cdot\left[ (\mbox{div.terms})
                + (a_4+a_4'\log\frac{\Lambda_M}{g_3^2}) g_3^6 \right.\\
            & & \left. + (\mbox{higher ord.}) \right]
\end{eqnarray*}

The first terms are divergent with some power of $\Lambda_M$, but they cancel
against analogous terms in the electric sector.
So the numbers $a_4$ and $a_4'$ carry the interesting finite information 
and are to be determined in lattice calculations.

On the lattice it is easier to use the internal energy $\epsilon$
\begin{eqnarray*}
  \epsilon   &   =  & \frac{T^2}{V} \frac{d}{dT} \log Z
                      = - T^2 \frac{dg_3^2}{dT} \epsilon_3 \\
  \epsilon_3 &\equiv& - \frac{1}{V} \frac{d}{dg_3^2} \log Z \\
             &   =  & - 3 \Lambda_M^3 \langle P \rangle \frac{\beta_3}{g_3^2}
                      \qquad , \beta_3 = \frac{2N}{a g_3^2}
\end{eqnarray*}
because it is related to the Plaquette expectation value $\langle P \rangle$.
We expand $\langle P \rangle$ to identify the relevant term ($\sim
\beta_3^{-4}$) using the ansatz
$\langle P \rangle = \sum\limits_{n=1} c_n \beta^{-n}$
where the coefficients $c_1$ and $c_2$ are known from lattice perturbation
theory. Further we define
\begin{displaymath}
  \Delta = \beta_3^3 ( \langle P \rangle - c_1 \beta_3^{-1}
           -c_2 \beta_3^{-2} ).
\end{displaymath}

Simulations were performed in 3 dimensional SU(2) and SU(3) on $16 \times 16
\times 64$ resp. $32^3$ lattices with Plaquette measurements for
$\beta_3\in[6.0;14.0]$ resp. $\beta_3\in[12.0;30.0]$.
The results for $\Delta$ are shown in figures 1 and 2.

\begin{figure}[htbp]
  \epsfig{file=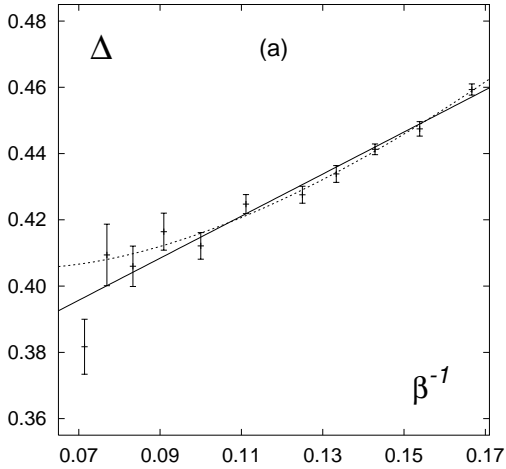,              width=70mm}
  \vspace*{-8.0ex}
  \caption{$\Delta$ versus $1/\beta_3$ in SU(2). The solid (dashed) line is the
    case I (case II) fit both with $c_5=0$.}
\end{figure}

\begin{figure}[htbp]
  \epsfig{file=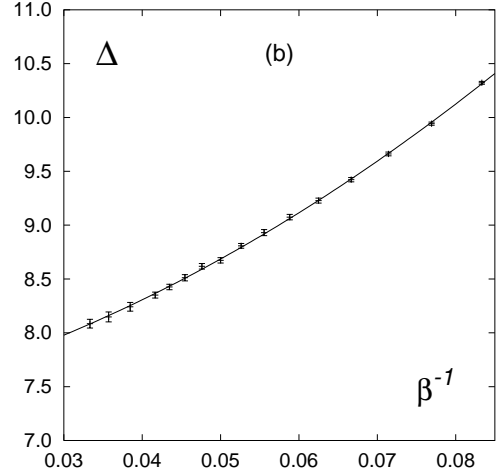,              width=70mm}
  \vspace*{-8.0ex}
  \caption{$\Delta$ versus $1/\beta_3$ in SU(3). The curve shows the case I and
    case II ($c_5=0$) fits, which lie on top of each other.}
\end{figure}

\noindent
With the fit-functions 
\begin{displaymath}
  \Delta = \left\{
  \begin{array}[c]{ll}
    c_3 + \frac{c_4}{\beta_3} + \frac{c_5}{\beta_3^2} & , \mbox{case I} \\
    c_3 + (c_4' \log \beta_3 + c_4) \frac{1}{\beta_3}
    + \frac{c_5}{\beta_3^2} & , \mbox{case II}
  \end{array} \right.
\end{displaymath}

\noindent
we got the following parameters:

\begin{tabular}{|c|c|c|}
  \hline
  SU(2)  & case I    &  case II   \\
  \hline
  $c_3$  & 0.351(5)  & 0.45(5)    \\
  $c_4$  & 0.635(37) & 1.4(4)     \\
  $c_4'$ & --        & -0.75(35)  \\
  $c_5$  & --        & --         \\
  $a_4$  & -0.010(1) & -0.002(16) \\
  \hline\hline
  SU(3)  & case I    &  case II   \\
  \hline
  $c_3$  & 7.30(11)  & 8.11(20)   \\
  $c_4$  & 15.2(3.6) & 99(6)      \\
  $c_4'$ & --        & -29.2(3.4) \\
  $c_5$  & 252(29)   & --         \\
  $a_4$  & -0.07(2)  & -0.17(6)   \\
  \hline
\end{tabular}

\vspace*{2.0ex}
In case II it was not possible to fit the logarithmic correction and
the quadratic term at the same time, so we set $c_5=0$. Although the
SU(2) data indicate a quadratic behaviour the case I fits were unstable
with a free $c_5$.

From the coefficients $c_i$ one can then extract $a_4$ by comparing the
expansions of $\epsilon_3$ and $f_M^{(0)}$.


\section{Conclusions, Part A}
The analysis presented above gives the framework to deternine the ${\cal
  O}(g^6)$ contribution to the free energy from lattice input.
At present the data are not accurate enough to be sensitive for logarithmic
corrections. 
A calculation of $c_3$ in lattice perturbation theory, which is in principle
possible, would, however, improve the analysis.
The ${\cal O}(g^6)$ contribution is about 10\% of the Stefan-Boltzmann value
for $g^2\simeq1$, which is in magnitude compatible with the known perturbative
contributions.

The contribution of the magnetic sector can be explained with the
existence of a pseudo particles with mass $m_M \sim g^2 T$
and degeneracy $N_D = 2(N^2-1)$. It would give the following contribution to
the energy density:
\begin{displaymath}
  \epsilon = T^2\,\frac{dg_3^2(T)}{dT}
  \cdot \frac{N_D m_M^3}{4 \pi g_3^2} .
\end{displaymath}

Using our fit results we obtain the following gluon masses:
\begin{displaymath}
  m_{gluon} = \left\{
  \begin{array}[c]{ll}
    (0.397 \pm 0.008)~g_3^2 &, \mbox{SU(2), I} \\
    (0.55 \pm 0.04)~g_3^2   &, \mbox{SU(3), I} \\
    (0.78\pm 0.29)~g_3^2    &, \mbox{SU(3), II}
  \end{array} \right.
\end{displaymath}
This may be compared with calculations of the magnetic mass from the gluon
propagator in Landau gauge \cite{hel95}, $m_{gluon} = 0.371(27)~g_3^2$
for SU(2).

\section{Part B: Spatial String Tension}
The motivation to study 3-d string tension and 4-d spatial string tension is twofold:
First one wants to know how good the reduction to MQCD works. And the second
question is, to what extent is the non-vanishing spatial string tension related
to the magnetic mass?
This determination of $\sigma_s$ has already been done for SU(2) and is now
repeated for SU(3).

In 4 dimensions a pseudo potential can be extracted from space-like Wilson
loops. Also above $T_c$ this has a non vanishing linear coefficient,
$\sigma_s$.
We analysed 10 $\beta$ values between 6.1 ($1.1\,T_c$) and 7.2 ($4.5\,T_c$)
on a $32^3\times 8$ lattice. More details about the determination of
$\sigma_s$ can be found in \cite{eos96}.

At high temperature an identification with $\sigma_3$ from MQCD is natural.
\begin{displaymath}
  \sqrt{\sigma_3} \sim g_3^2 \quad, \quad g_3^2 = g^2(T) T
                             \quad, \quad \sqrt{\sigma_s} \sim g^2(T) T
\end{displaymath}

A 2-loop fit yields $\sqrt{\sigma_s(T)} = c \cdot g^2(T) \cdot T$ with
$c = 0.566(13)$. The spatial string tensions and the fit are shown in
figure 3.

From a corresponding analysis in the three dimensional SU(3) theory
\cite{lue94} we get $\sqrt{\sigma_3} = 0.544(4) \cdot g_3^2$.

\begin{figure}[htbp]
  \epsfig{file=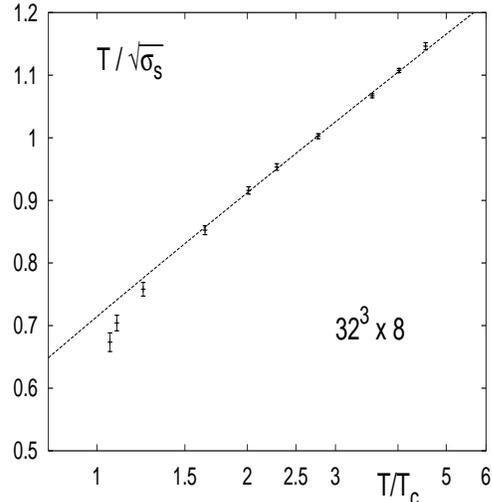, height=70mm, width=70mm}
  \vspace*{-8.0ex}
  \caption{$T/\sqrt{\sigma_s(T)}$ vs. $T/T_c$ for SU(3)}
\end{figure}

\section{Conclusions, Part B}
Our data confirm a $g^2(T) T$ behaviour of $\sqrt{\sigma_s}$ above $1.5 T_c$.
There is a 90 \% coincidence with the numerical value for $\sqrt{\sigma_3}$
from MQCD.
The numbers are also very similar to the estimated magnetic masses from part A,
$m_{gluon} \approx (0.5 - 0.8) \cdot g_3^2$.

\end{document}